\documentclass[conference]{IEEEtran}
\IEEEoverridecommandlockouts
\usepackage{cite}
\usepackage{amsmath,amssymb,amsfonts}
\usepackage{algorithmic}
\usepackage{graphicx}
\usepackage{textcomp}
\usepackage{xcolor}
\usepackage[inline]{enumitem}
\setlist[itemize]{leftmargin=*}
\usepackage{graphicx}
\usepackage{subfigure}
\usepackage{multirow}
\usepackage{booktabs}
\usepackage{pifont}
\usepackage{bm}

\def\BibTeX{{\rm B\kern-.05em{\sc i\kern-.025em b}\kern-.08em
    T\kern-.1667em\lower.7ex\hbox{E}\kern-.125emX}}
\begin{document}

\title{ChartAdapter: Large Vision-Language Model for Chart Summarization
}

% \author{\IEEEauthorblockN{Anonymous Authors}}
\author{\IEEEauthorblockN{Peixin Xu}
\IEEEauthorblockA{\textit{The Hong Kong Polytechnic University} \\
Hong Kong SAR\\
james-x.xu@connect.polyu.hk}
\and
\IEEEauthorblockN{Yujuan Ding}
\IEEEauthorblockA{\textit{The Hong Kong Polytechnic University} \\
Hong Kong SAR\\
dingyujuan385@gmail.com}
\and
\IEEEauthorblockN{Wenqi Fan}
\IEEEauthorblockA{\textit{The Hong Kong Polytechnic University} \\
Hong Kong SAR\\
wenqifan03@gmail.com}
}

\maketitle

\begin{abstract}
Chart summarization, which focuses on extracting key information from charts and interpreting it in natural language, is crucial for generating and delivering insights through effective and accessible data analysis. Traditional methods for chart understanding and summarization often rely on multi-stage pipelines, which may produce suboptimal semantic alignment between visual and textual information. In comparison, recently developed LLM-based methods are more dependent on the capability of foundation images or languages, while ignoring the characteristics of chart data and its relevant challenges. To address these limitations, we propose ChartAdapter, a novel lightweight transformer module designed to bridge the gap between charts and textual summaries. ChartAdapter employs learnable query vectors to extract implicit semantics from chart data and incorporates a cross-modal alignment projector to enhance vision-to-language generative learning. By integrating ChartAdapter with an LLM, we enable end-to-end training and efficient chart summarization. To further enhance the training, we introduce a three-stage hierarchical training procedure and develop a large-scale dataset specifically curated for chart summarization, comprising 190,618 samples. Experimental results on the standard Chart-to-Text testing set demonstrate that our approach significantly outperforms existing methods, including state-of-the-art models, in generating high-quality chart summaries. Ablation studies further validate the effectiveness of key components in ChartAdapter. This work highlights the potential of tailored LLM-based approaches to advance chart understanding and sets a strong foundation for future research in this area.

% Recently Multi-modal Large Language Models (MLLMs) or Large Visual-Language Models (LVLMs) have been the hottest topics in academia. Analysis description generation takes advantage of the strong representation and capturing abilities of MLLM, generating fluent, native, and accurate summarization. 
% Based on MLLM-based approaches, we proposed a \textbf{\underline{Text Component}} structure with a well-designed multi-stage training strategy, explicitly modeling different mindsets for various chart types. 
% Exhaustive experiments with the current SOTA of Chart Question Answering (CQA) and Visual Question Answering (VQA) along with several models in the Chart-to-text Generation field showed that our text component design improves generation qualities by extracting high-level semantic mindsets for different chart types. 
% Further ablation experiments verified the effectiveness of each structure in the design text component. 
\end{abstract}


\section{Introduction}
Charts are a powerful and widely used format for illustrating data, playing a critical role in various domains such as business, social sciences, and natural sciences~\cite{zhang2024tinychart,ding2024fashionregen}. They serve as a compact and effective medium for presenting complex data and enabling insights to be drawn efficiently. With the advent of large language models (LLMs), significant progress has been made in understanding and interpreting different data modalities, including text, images, and multimodal data~\cite{zhao2023survey,fan2024rag,fan2024graph}. However, charts—a unique multimodal data type combining textual and visual information—have been relatively underexplored in the context of LLM-based methods. This gap is particularly concerning, as understanding and interpreting charts are essential for downstream applications like data-driven decision-making, report generation, and knowledge dissemination. Despite the advances in general multimodal language models, the development of chart-specific models remains insufficient, with existing methods often yielding suboptimal performance~\cite{fu2022chartstamp, liu2023deplot}.

Charts present a unique set of challenges due to their complex and hybrid data format~\cite{zhang2024tinychart}. They integrate visual elements, such as bars, lines, or pie slices, with textual components, such as titles, labels, and legends~\cite{masry2022chartqa, zhang2024tinychart}. This dual nature makes charts highly informative but also challenging to process and interpret. Accurately summarizing chart content requires extracting and aligning semantic information from both the visual and textual modalities, which often exhibit varying levels of complexity and abstraction~\cite{Reiter2007anarchitectureb5}. The interplay between these modalities demands sophisticated techniques to bridge the gap, a need that existing methods often fail to meet effectively.

Traditional methods for chart summarization typically involve a multi-stage pipeline, where information is first extracted from charts and then used to generate textual summaries~\cite{shankar2022chart2text, raian2023chartsumm, panupong2015compositionalb8, fu2022chartstamp, liu2023deplot}. While functional, these approaches are not integrated, as the individual components are optimized independently, leading to inefficiencies and suboptimal performance. More recently, several LLM-based methods have been introduced, leveraging the powerful capabilities of LLMs to process and generate chart-related descriptions~\cite{Han2023ChartLlama, Liu2023MatCha, Masry2024ChartInstruct, Masry2023UniChart, zhang2024tinychart}. However, these methods face significant limitations: they rely heavily on pre-trained models and their basic image/text processing capabilities, often neglecting the semantic alignment between chart data and textual summaries. Furthermore, training such models is inherently challenging, requiring high-quality, large-scale datasets and effective training strategies, which are often lacking.

To address these limitations, we propose ChartAdapter, a novel approach designed to enhance the alignment between charts and their textual summaries. ChartAdapter is a lightweight transformer-based module that employs a set of learnable query vectors to extract implicit semantics relevant to charts, complemented by a cross-modal alignment projector. This module acts as a bridge between chart encoders and LLMs, facilitating vision-to-language generative learning by connecting ChartAdapter's output to an LLM. To ensure effective training, we introduce a four-stage hierarchical training procedure that updates different parts of the learnable parameters progressively. Additionally, we developed a large-scale dataset specifically for chart summarization, sourced from existing chart-related datasets, to support effective training.

Extensive experiments were conducted on the standard Chart-to-Text testing set~\cite{shankar2022chart2text} to evaluate our proposed method against existing approaches, including state-of-the-art models. The results demonstrate that ChartAdapter significantly enhances the quality of chart summaries, outperforming competitive baselines. Ablation studies further validate the effectiveness of the proposed technical components, highlighting the potential of our approach to advance the field of chart understanding and interpretation.

\section{Related Works}
% \input{contents/charts_chart2text_stages}

% \subsection{Chart-to-text Generation}

\noindent \textbf{Chart-to-text Generation}. Compared to general images, essential features of charts contribute to LVLM-based generation, such as \
\begin{enumerate*}
\item meaningless blanks and blocks~\cite{zhang2024tinychart}, 
\item hints, legends, and other textual elements~\cite{masry2022chartqa}
\item accurate information extraction and numerical calculation~\cite{Han2023ChartLlama, zhang2024tinychart}
\end{enumerate*}
These features resulted in an increasing need for high-quality summarization. Open-source datasets provide summaries either from human-written,  papers or articles~\cite{shankar2022chart2text, raian2023chartsumm, Shankar2022opencqa}, or generated from text generation models (such as LLM) or pre-set templates~\cite{Reiter2007anarchitectureb5, zhang2024tinychart}.
Template-based generation methods shared a \textit{data table fill-in cloze} format, leading to human bias, inflexibility, and weakness in shifting trends representation despite their general applicability~\cite{shankar2022chart2text}. Therefore, text generation models~\cite{fu2022chartstamp, liu2023deplot} finetuning on \textit{Data-Table-to-Text} summarization were adopted to generate chart summaries according to chart tableau data extracted from off-the-shelf image encoders. However, these approaches suffered from error accumulation and suboptimal performance constrained by independent optimization. 
Consequently, MLLM-based solutions~\cite{Han2023ChartLlama, Masry2024ChartInstruct} provided end-to-end training strategies, enhancing chart understanding capabilities via reasoning and characterizing abilities of the LLM backbone. Although outstanding progress was achieved, they still suffered from extensive parameter size, efficient input representation, and also limited computing and data resources~\cite{zhang2024tinychart}.

% \subsection{Multi-modal Large Language Models}

\noindent \textbf{Multi-modal Large Language Models}. With the flourishing of LLMs, cutting-edge research \cite{Liu2023llava, bin2024gallery,li2024empowering} invoke LLM inputs from various modalities encoders, due to their robust capabilities with well-developed positional embeddings and advanced attention mechanisms. In general, the commonly employed framework can be summarized as \textit{modality components} $\xrightarrow{}$ \textit{alignment network} $\xrightarrow{}$ \textit{LLM backbone} $\xrightarrow{}$ \textit{modality outputs}. 
Preliminary stage LVLMs~\cite{fu2022chartstamp, liu2023deplot} process image encoders initialized by off-the-shelf tools and trained with frozen LLM backbones, while subsequent works~\cite{Wang2024qwen2vl, Liu2023llava} benefit from end-to-end training, enhancing a multi-modal embedding space rather than the fixed textual space. Those embeddings were injected into linguistic inputs according to a special token \textit{$<$image$>$} or \textit{$<$video$>$}. 
Because of the minor impact of placing the special token~\cite{Brandon2024mm1}, pioneer studies~\cite{Wang2024qwen2vl, Xue2024blip3} supported multiple images or clipping frames inputs.

% \subsection{Cross Modality Alignment} \label{lr:cross_modality_alignment}

\noindent  \textbf{Cross Modality Alignment}. The alignment network takes a tiny percentage but contributes a significant impact on the overall performance \cite{xu2024align,Wang2024qwen2vl}. Hence, numerous researchers devoted themselves to perfecting this structure, including MLP-based \cite{Liu2023llava, Liu2024Improvedb44, Wang2024qwen2vl}, Attention-based \cite{bin2024leverage}, and Learnable-parameters-based connectors \cite{Li2023blip2, Zhang2024llama_adapter}. 
Generally, the alignment network shoulder projection between different pre-trained vector spaces. This strong physical meaning led to the widespread adoption of the Multi-stage Training Strategy \cite{Bai2023qwenvl, Wang2024qwen2vl, Liu2023llava}. 
Jointly trained with image encoders in the beginning, alignment networks were trained with the LLM backbone to achieve better performance after the overall joint training phase. 
Apart from that, CLIP \cite{Alec2021clip} proposed a contrastive pre-training strategy to project similar representatives between different modal features of oneself while distinguishing from others.

\section{Methodology}
\begin{figure*}
    \centering
    \includegraphics[width=0.75\linewidth]{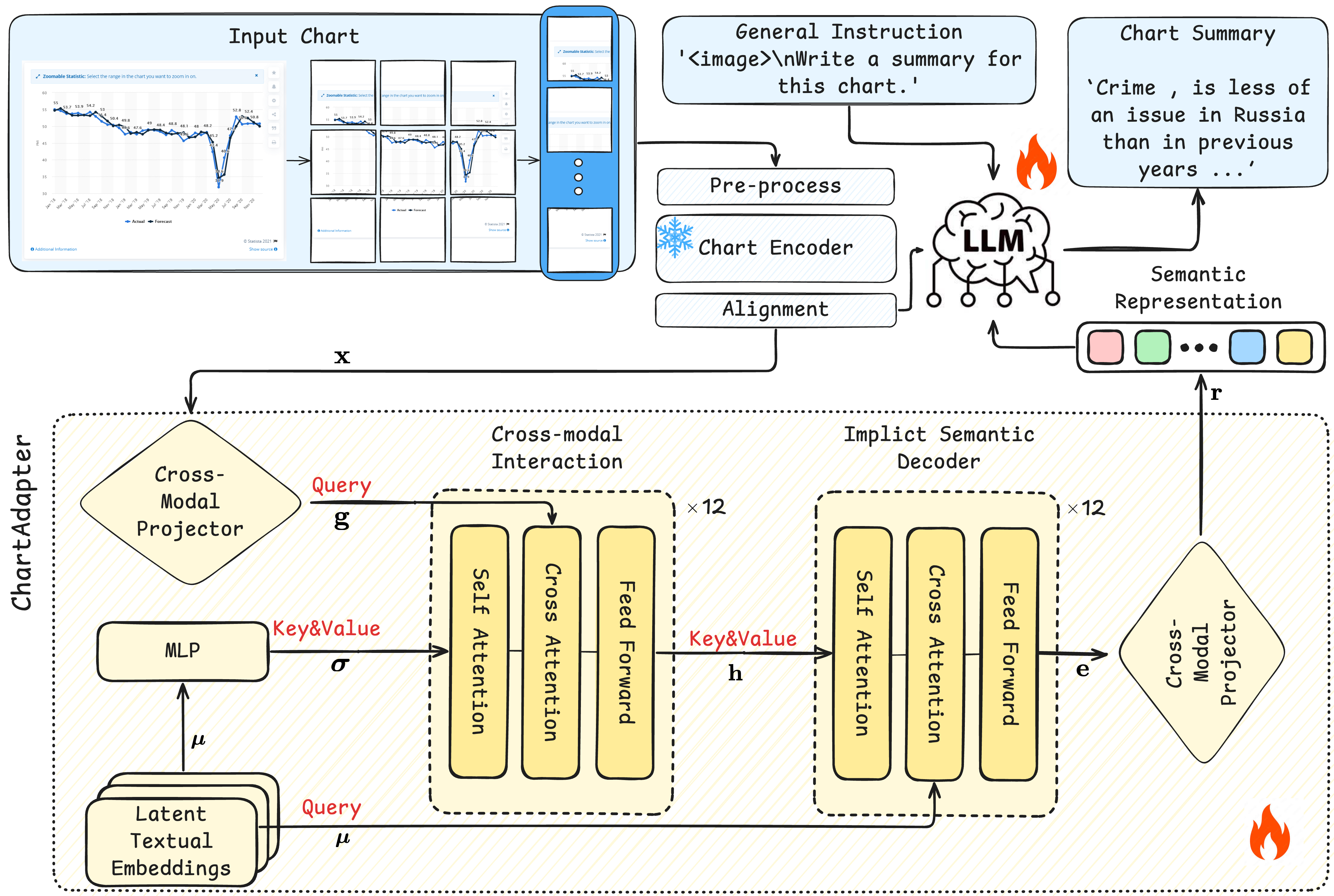}
    \caption{ The overall framework of the proposed \textbf{ChartAdapter} integrated in a Large Vision-Language Model (LVLM). It acts as a bridge between the chart encoder and LLM through transformer-based cross-modal interaction modeling with learnable latent embeddings to extract implicit semantics relevant to charts, complemented by a cross-modal alignment projector. }
    \label{fig:text_component_design}
\end{figure*}

% \subsection{Model Structure}
% Our ChartAdapter aims to extract different latent semantic representations for each chart, which aids in producing accurate descriptions and fluent summarization.  To further illustrate our design, the ChartAdapter has 4 components, including 
% \begin{enumerate*}[label=\textit{(\alph*)}]
% 	\item latent textual embeddings, 
% 	\item a shared projector, 
% 	\item a cross-modal interaction network, and 
% 	\item an implicit semantic decoder
% \end{enumerate*}.
% Figure \ref{fig:text_component_design} shows both the detailed structure of our ChartAdapter and the overall pipeline.

In this paper, we propose ChartAdapter, a novel framework for chart summarization. Our method employs a large language model (LLM) as the backbone for its robust textual understanding and generation capabilities, complemented by a visual encoder to provide foundational chart understanding. The core of our approach lies in the integration of these two components through the ChartAdapter module, which enhances the synergy between chart understanding and summary generation. 

\subsection{ChartAdapter}
ChartAdapter serves as a bridge between the visual encoder and the language model, designed to facilitate efficient and effective chart summarization. Structurally, it consists of four key components: the cross-modal projector, latent textual embeddings, the cross-modal interaction layer, and the implicit semantic decoder layer. These components work together to align and unify multimodal embeddings, extract relevant visual information, facilitate cross-modal interactions, and decode visual semantics into textual summaries.

The modular design of ChartAdapter ensures flexibility and scalability, allowing seamless integration with both the upstream visual encoder and the downstream language model. This hierarchical structure enables effective collaboration between the chart understanding module and the summary generation backbone.

\subsubsection{Cross-Modal Projector}
The cross-modal projector acts as a critical mapping layer with two primary functions: (1) \textbf{Embedding Alignment}: Through effective training, the projector aligns the embedding spaces of the backbone LLM and the lightweight ChartAdapter module, which is initialized with BERT embeddings for computational efficiency; and (2) \textbf{Dimensional Unification}: The projector unifies the dimensions of the visual encoder and the language model, facilitating seamless integration between upstream chart processing and downstream textual generation modules. We specifically apply a transaction matrix on the extracted chart features $\bm{x}$ to function as the projector as follows:
\begin{equation}
    \bm{g} = \mathcal{G}(\bm{x}) = \mathbb{M}^{d_v\times d_t}\bm{x},
\end{equation}
%%% unpolished %%%
where $d_v$ and $d_t$ are the dimensions of the visual feature from the Chart Encoder and that of the textual feature operated in ChartAdapter. 
% $\mathcal{G}(\cdot)$ refers to the projector network, that we implemented as a projection matrix $\mathbb{M} \in \mathbb{R}^{c_{out}, c_{in}}$, where $c_{out}$ and $c_{in}$ are embedding sizes of the LVLM and the ChartAdapter. 

\subsubsection{Latent Textual Embeddings}
Latent textual embeddings $\bm{\mu}$ serve as cross-modal gap bridges, extracting the most relevant and significant visual information from charts for summarization. These embeddings are a set of learnable parameters designed to encode latent semantics present in the input charts. They can be further decoded to generate textual summaries, capturing key insights from the visual data with a two-layer Multi-Layer Perception (MLP) module as follows:
\begin{equation}
\begin{split}
    \bm{\sigma} = ReLU(w_1 \cdot (ReLU(w_2 \cdot \bm{\mu} + b_2)) + b_1), 
\end{split}
\end{equation}
where $w_1$, $w_2$, $b_1$, $b_2$ are weights and bias of the MLP module. 
% $\mathcal{F}(\cdot)$ refers to the (?) network between latent textual embeddings $\bm{\mu}$ and latent semantic angles $\bm{\sigma}$, which was two MLP layers with ReLU as the activate function in implementation. 

\subsubsection{Cross-Modal Interaction Layer}
This layer is implemented as a transformer-based cross-attention mechanism that integrates vision-derived features into the LLM’s text prediction process. By fostering interaction between the chart’s visual representations and the language model’s textual capabilities, the cross-modal interaction layer enhances the overall coherence and relevance of the generated summaries. It specifically operates as follows:
% \begin{equation}
%     \bm{\sigma_{i}} = \frac{\mathcal{SF}(W_q^i \bm{\sigma} \times {W_k^i \bm{\sigma}} ^T)}{\bm{d_{\bm{\sigma}}}} W_v^i \bm{\sigma}
% \end{equation}
% \begin{equation}
% \begin{split}
%     \bm{m}_{i+1} & = \frac{\mathcal{SF}(\bm{m}_i \times {\bm{\sigma^{i}}} ^T)}{d_{\bm{\sigma^{i}}}} \bm{\sigma_{i}}\\
%     & where\ \bm{m}_0 = \bm{g}
% \end{split}
% \end{equation}
\begin{equation}
    \bm{h} = \text{Attention}(\bm{g}, \bm{\sigma}, \bm{\sigma}). 
\end{equation}
%%% unpolished %%%
% $\mathcal{SF}(\cdot)$ refers to the softmax function, $\bm{\sigma_{i}}$ means latent semantic angles after self-attention in each layer, $\bm{m}_i$ means the fused multi-modal features as cross-modal interaction network's output. We labeled the output from the last layer as $\bm{m}$

\subsubsection{Implicit Semantic Decoder Layer}
The implicit semantic decoder layer, also built upon a cross-attention mechanism, interprets the visual semantic representations to guide the final textual summary generation. This layer operates after an inverse cross-modal projection procedure, ensuring that the extracted visual features are effectively transformed into meaningful textual outputs. 
\begin{equation}
    \bm{e} = \text{Attention}(\bm{\mu}, \bm{h}, \bm{h}). 
\end{equation}
% \begin{equation}
%     \bm{n}_{i}  = \frac{\mathcal{SF}(W_q^i \bm{m} \times {W_k^i \bm{m}} ^T)}{\bm{d_{\bm{m}}}} W_v^i \bm{m}
% \end{equation}
% \begin{equation}
% \begin{split}
%     \bm{l}_{i+1} & = \frac{\mathcal{SF}(\bm{l}_i \times {\bm{n}_{i}} ^T)}{\bm{d_{\bm{n}_{i}}}} \bm{n}_{i}\\
%     & where\ \bm{l}_0 = \bm{\mu}
% \end{split}
% \end{equation}
%%% unpolished %%%
% $\bm{n}_{i}$ means the fused feature after self-attention module in each layer, $\bm{l}_i$ means the latent representations in ChartAdapter space generated from each layer. We take the last one as $\bm{l}$ and project it into the LVLM space. 
The fused feature after the last layer of Implicit Semantic Decoder is further projected back to the visual space with the transposed projector as the semantic representation to facilitate the summary generation for the LLM as
\begin{equation}
    \bm{r} = \mathbb{M}^{T} \bm{e}. 
\end{equation}
% %%% unpolished %%%
% $\bm{e}$ refers to the semantic representations as our ChartAdapter's output. 
The composition of different modules of ChartAdapter is shown in Fig.~\ref{fig:text_component_design}. In the final step, the framework utilizes the LLM to generate the chart summary $s$ by leveraging the outputs of the Chart Encoder $\bm{g}$ and ChartAdapter $\bm{r}$, combined with the general textual instructions $p$ as follows: 
\begin{equation}
    s = \text{LLM}(p, \bm{r}, \bm{g}). 
\end{equation}

\subsection{Training Strategy}
The training of our chart summarization model, ChartAdapter, follows a three-stage hierarchical learning strategy, progressively optimizing the model parameters from local to global levels. In the \textbf{first stage}, we focus on aligning the embedding spaces between ChartAdapter and the chart summarization backbone (an LLM) by optimizing the cross-modal projector using a contrastive learning approach. This step ensures effective alignment of visual and textual representations. In the \textbf{second stage}, we optimize the entire ChartAdapter component, updating all its trainable parameters, including the projector, multi-layer perceptron (MLP), latent textual embedding, and two attention modules. Finally, in the \textbf{third stage}, we fine-tune the LLM in conjunction with ChartAdapter on the chart summarization task, enabling the integrated system to generate high-quality summaries. This progressive strategy ensures robust feature alignment and enhanced task performance.

\section{Experiments}
\subsection{Experimental Settings}

\subsubsection{Dataset}
After surveying existing chart-related datasets, we found the limitations of existing datasets for chart summarization. Even though there are several datasets available and applicable to the targeted task, none of them are sufficiently robust and comprehensive for the specific investigation of the chart summarization task, particularly in the context of training or fine-tuning large language models. For instance, ChartSumm~\cite{raian2023chartsumm}, the most relevant dataset, contains only 70,000+ samples, limiting its scale and diversity. To overcome these challenges, we developed a new dataset \textit{\textbf{ChartSumm}} specifically for chart summarization, sourced from existing datasets including ChartSumm~\cite{raian2023chartsumm}, Chart-to-Text~\cite{shankar2022chart2text}, OpenCQA~\cite{Shankar2022opencqa}, TinyChart~\cite{zhang2024tinychart}, and others. Our dataset is large-scale,  comprising a total of 190,618 samples, as detailed in Table 1. It is more comprehensive in terms of source variety, chart types, and topic coverage, while maintaining a balanced distribution of complexity across different chart difficulties. This ensures a more effective and reliable foundation for LLM-based chart summarization tasks. Experiments, we sampled 1,000 examples from this dataset to construct a validation set, leaving the remaining 189,618 samples for training. For testing, we used the standard testing set, chart2text pew test set~\cite{shankar2022chart2text}, which is commonly applied in prior works to ensure fair comparisons between our method and baseline approaches. Please refer to Table.~\ref{Table:training_data_description} for more details of the dataset.

\begin{table}[ht]
    \centering
    \begin{tabular}{c|c|c|c}
    \toprule
        \multicolumn{2}{c|}{\textbf{Dataset}} &\textbf{Chart Typ}e & \textbf{Sample Num} \\
        \midrule
        \multicolumn{2}{c|}{Chart2Text8k \cite{shankar2022chart2text}} & mixed & 7862 \\
        \hline
		\multirow{4}*{Chart2Text \cite{shankar2022chart2text}} & \multirow{2}*{Pew} & complex & 26476 \\
		\cline{3-4}
		 &  & simple & 5092 \\
		\cline{2-4}
		 & \multirow{2}*{Statista} & complex & 11774 \\
		\cline{3-4}
		 &  & simple & 47404 \\
		\hline
        \multicolumn{2}{c|}{ChartSumm \cite{raian2023chartsumm}} & simple & 75251 \\
        \hline
        \multicolumn{2}{c|}{OpenCQA \cite{Shankar2022opencqa}} & mixed & 5588 \\
        \hline
        \multicolumn{2}{c|}{VisText \cite{Benny2023vistext}} & simple & 11171 \\
        \hline
        \multicolumn{3}{c|}{\textbf{Total}} & 190618 \\ \midrule
    \end{tabular}
    \caption{Statistics of the \textbf{\textit{ChartSumm}} Dataset}
    \label{Table:training_data_description}
\end{table}

\subsubsection{Baselines} 
After surveying previous chart-to-text generation approaches, we selected the following models from six categories for comparison, including,
\begin{enumerate}[label=\textit{(\alph*)}]
    \item \textbf{OCR-based models}: \textit{OCR-Field\_infuse} (short in OCR-Field), \textsc{\textit{OCR-BART}} and \textsc{\textit{OCR-T5}}~\cite{raian2023chartsumm}. 
    \item \textbf{Off-the-shelf LVLM-based method}: LLaVa 1.5~\cite{Liu2023llava}. 
    \item \textbf{Early End-to-end training methods}: \textsc{\textit{Pix2struct}}~\cite{Kenton2023pix2struct} and \textsc{\textit{Unichart}}~\cite{Masry2023UniChart}. 
    \item \textbf{Fine-tuned LVLM-based methods}: \textsc{\textit{Matcha}}~\cite{Liu2023MatCha}, \textsc{\textit{ChartInstruct}}~\cite{Masry2024ChartInstruct}, \textsc{\textit{ChartLlama}}~\cite{Han2023ChartLlama} and \textsc{\textit{ChartAst}}~\cite{Meng2024ChartAssisstant}. 
    \item \textbf{State-of-the-art chart-to-text model}: TinyChart~\cite{zhang2024tinychart}. 
    \item \textbf{State-of-the-art vision-to-text model}: Qwen2VL~\cite{Wang2024qwen2vl}. 
\end{enumerate}

\subsubsection{Model settings}
In our experiment, we adopted Qwen2VL-2B~\cite{Wang2024qwen2vl} as the backbone, including its LLM, image encoder part and its token merging strategy. The cross-modal interaction layers and the implicit semantic decoder in ChartAdapter were initialized by Bert-base-uncase~\cite{Jacob2018BERT} for its outstanding text sequence processing ability and good efficiency. We implement Attention in ChartAdapter with a 12-layer multi-head attention structure. 
% In this experimental group, we implemented our three-stage hierarchical training process. We label this experimental group as \textit{\textbf{ChartAdapter}}, which is trained on our three-stage hierarchical training process. 

\begin{table}[ht]
    \centering
    \begin{tabular}{l|c|c|c c c}
    \toprule
         Model Name & Size & BLEU-4 & Rouge-1 & Rouge-2 & Rouge-L \\ \midrule
         OCR-Field~\cite{shankar2022chart2text}  & 1.5b & 0.19 & - & - & - \\
         OCR-BART~\cite{shankar2022chart2text}  & 0.5b & 9.09 & - & - & - \\ 
         OCR-T5~\cite{shankar2022chart2text}  & 0.7b & 10.49 & - & - & - \\ \midrule
         Pix2struct~\cite{Kenton2023pix2struct} & 0.3b & 10.30 & - & - & - \\ 
         Unichart~\cite{Masry2023UniChart}  & 1b & 12.48 & - & - & - \\ \midrule
         LlaVa1.5~\cite{Liu2023llava}  & 13b & 7.16 & - & - & - \\
         Qwen2VL~\cite{Wang2024qwen2vl}  & 2b & 28.73 & 34.14 & 11.57 & 21.33 \\ 
         Qwen2VL~\cite{Wang2024qwen2vl}  & 7b & 30.16 & 37.07 & 13.72 & 22.84 \\ \midrule
         Matcha~\cite{Liu2023MatCha}  & 8b & 12.20 & - & - & - \\ 
         ChartInstruct~\cite{Masry2024ChartInstruct}  & 7b & 13.83 & - & - & - \\
         ChartLlama~\cite{Han2023ChartLlama}  & 13b & 14.23 & - & - & - \\
         TinyChart~\cite{zhang2024tinychart}  & 3b & 17.18 & - & - & - \\ \midrule
         ChartAdapter & 3b & \textbf{\underline{35.55}} & \textbf{\underline{41.49}} & \textbf{\underline{15.75}} & \textbf{\underline{25.79}} \\ \midrule
    \end{tabular}
    \caption{Model Performances on Chart-to-text Pew Dataset}
    \label{Table:model_performance}
\end{table}

\subsubsection{Evaluation settings}
% Chart summarization performances published from these approaches build up a comprehensive perspective of existing chart-to-text methods, including state-of-the-art models on the standard Chart-to-Text testing set~\cite{shankar2022chart2text}. 
In line with previous chart-to-text research~\cite{shankar2022chart2text}, we evaluate these performances on the BLEU-4 metric, the widely used standard, along with Rouge-1, Rouge-2, and Rouge-L. These metrics embodied the fluency, similarity, and accuracy of the generated chart summarization. 
% We further evaluate our first-stage training on the cross-modal projector with the cross-modal matching task.
% Due to our hierarchical training process, the objective in stage 1, the contrastive learning stage, is different from others. We additionally evaluate its effect by analyzing the binary classification ability of the projector network, labeling the similarity score to itself as 1 while 0 to others. 

\subsection{Overall Performance of ChartAdapter}
As illustrated in Table~\ref{Table:model_performance}, overall, our \textit{\textbf{ChartAdapter}} (BLEU-4: 35.55, Rouge-1: 41.49, Rouge-2: 15.57, Rough-L: 25.79) generates summaries with more accurate wordings, more fluent sentences, and more precise descriptions compared to existing methods, even to the SOTA model TinyChart with BLEU-4 scores 17.18. Compared to other fine-tuned LVLM-based methods, our ChartAdapter has a 21.2\% improvement in BLEU-4.
Our ChartAdapter design enhanced the Qwen2VL model with outperformed capabilities to an even larger version. 
Efficient improvements have been accomplished through our ChartAdapter implementation, with a slight scaling (parameter size from 2 billion to 3 billion) achieving significant improvement to the backbone even with a larger scale. 
Improvement contains a 17.87\% / 23.74\% relevant improvement in BLEU-4 and an average relevant improvement of 13.21\% / 26.19\% in Rough-related metrics compared to Qwen2VL-2B~\cite{Wang2024qwen2vl} or Qwen2VL-7B~\cite{Wang2024qwen2vl}, indicating a robust chart summarization ability. More details can be referred to Table.~\ref{Table:model_performance}.

\subsection{Ablation Study of ChartAdapter}
Several ablation experiments are conducted to fully verify our ChartAdapter design and training strategy. Specifically, we investigate the following version of ChartAdapter as shown in Table.~\ref{Table:exp_control_groups_definition}. The performance of these ChartAdapter variants is illustrated in Table~\ref{Table:model_ablation_experiment}

\begin{table}[ht]
    \centering
    \begin{tabular}{l|c c c}
    \toprule
        Group Name & Stg.1 & Stg.2 & Stg.3  \\ \midrule
        ChartAdapter & \checkmark & \checkmark & \checkmark  \\ 
        ChartAdapter w/o Stg.1 & \ding{55} & \checkmark & \checkmark  \\
        ChartAdapter w/o Stg.2 & \checkmark & \ding{55} & \checkmark  \\
        ChartAdapter ChA only & \checkmark & ChA & ChA  \\
        ChartAdapter LLM only & \checkmark & LLM & LLM  \\ \midrule
    \end{tabular}
    \caption{Ablated models of ChartAdapter}
    \label{Table:exp_control_groups_definition}
    \footnotesize{Stg.1/2/3 refers to stage 1/2/3, \checkmark means we conducted this stage while \ding{55} means we didn't, ChA means we only update ChartAdapter, LLM means we only update Latent Textual Embedding and the LLM backbone.}
\end{table}

We first evaluate the effect of the first stage of cross-modal projector training for embedding space alignment. By comparing the results of \textit{ChartAdapter w/o Stg.1} and \textit{ChartAdapter}, we can observe that without the first stage of fine-tuning the projector, the overall performance for Chart Summarization fluctuates under 2\% with higher initial losses, indicating good initialization which only accelerates the convergence process. 

% \input{contents/results/table_contrastive_learning}

% Stage 1 enhanced the cross-modal projector with robust discrimination when projecting linguistic embeddings from one to another space. Our experimental results suggest that the learned projectors achieve 0.98 in AUC for cross-modal matching  indicates that projected vectors had an outstanding consistency since they were more similar to themselves while distinguishable to others. Other binary classification metrics also shown a satisfying result, with 71\% in accuracy, 72\% in precision, 71\% in recall, and 71\% in f1-score. The contrastive loss enhanced the projector network with proper initialization. 
% with  stage, we implemented an ablation group with the same settings as the experimental group, except for the training strategy. In this group, the projector network of ChartAdapter is randomly initialized and directly starts from the cross-modal summarization learning stage. We labeled this group as \textit{\textbf{Ours-Qwen2VL w/o clip}}. 

We further evaluate the effect of the second-stage learning, \textit{i.e.}, solely optimizing parameters of ChartAdapter. As we can see from the result of \textit{ChartAdapter w/o Stg.2}, there is a performance drop comparing to \textit{ChartAdapter}, which is also more significant than that of \textit{ChartAdapter w/o Stg.1}.
% cross-modal summarization learning stage, we implemented an ablation group without the contrastive learning stage. This group used the projector network trained by the contrastive learning stage to jointly optimize the ChartAdapter and the LLM backbone. We labeled this group as \textit{\textbf{Ours-Qwen2VL single}}. 

Finally, we evaluate the different settings of parameter updating in stage 3, specifically, solely fine-tuning the ChartAdapter(ChA) part or the LLM part. From the results we can see our setting of joint learning both parts play a vital role to the overall performance as both the ablated models perform badly according to the performance, especially the version of tuning the LLM part only, which achieves the worst performance among all compared ablation models. 
\begin{table}[ht]
    \centering
    \begin{tabular}{l|c|c c c}
    \toprule
         Model Name  & BLEU-4 & Rouge-1 & Rouge-2 & Rouge-L \\ \midrule
         ChartAdapter & \textbf{\underline{35.55}} & \textbf{\underline{41.49}} & \textbf{\underline{15.75}} & \textbf{\underline{25.79}} \\ 
         ChartAdapter  w/o Stg.1  & \textbf{\underline{35.58}} & \textbf{\underline{40.93}} & \textbf{\underline{15.37}} & \textbf{\underline{25.33}} \\
         ChartAdapter  w/o Stg.2  & 34.48 & 40.15 & 15.02 & 25.01 \\
         ChartAdapter ChA only  & 28.83 & 35.87 & 12.68 & 22.62 \\
         ChartAdapter LLM only  & 8.63 & 9.72 & 6.37 & 8.09  \\ \midrule
    \end{tabular}
    \caption{Ablation Study Results}
    \label{Table:model_ablation_experiment}
\end{table}
\begin{figure}
    \centering
    \includegraphics[width=1\linewidth]{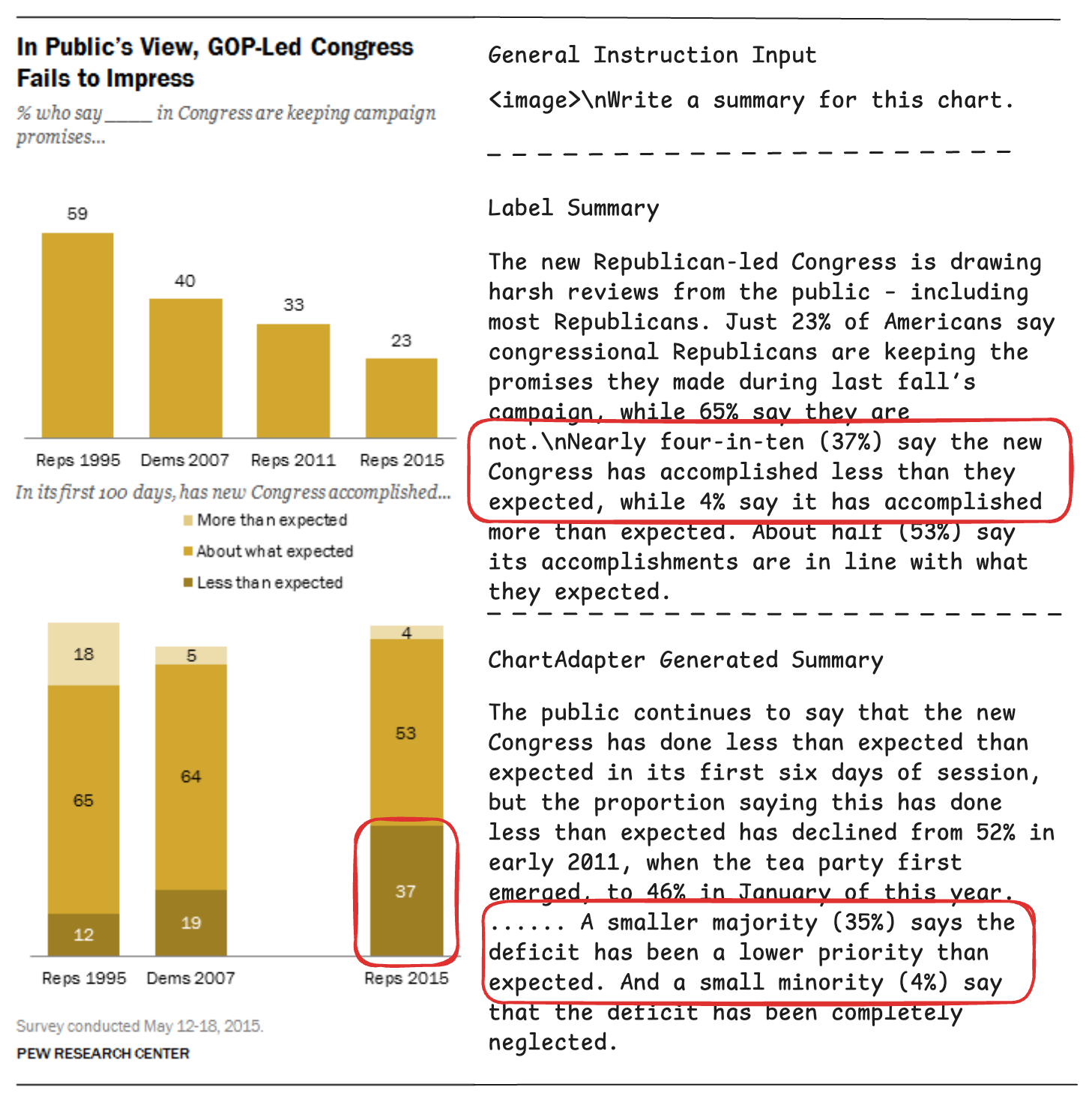}
    \caption{Chart Summarization Sample Generated by ChartAdapter. }
    \label{fig:case}
\end{figure}
\subsection{Case Illustration}
We illustrate a chart summarization case in Fig.~\ref{fig:case}. In this case, we demonstrate the capabilities of our chart summarization model, which enables the model to generate natural and fluent summaries that effectively describe the key points in complex charts. Furthermore, The model not only captures the core insights but also extracts trend information with remarkable precision, providing a clear and coherent narrative. By automatically translating the visual data into descriptive text, it highlights patterns, changes, and correlations in the data, showcasing the effectiveness of our model in making chart information more accessible and understandable. This case underscores the model's ability to transform data visualization into insightful, human-readable summaries, offering a powerful tool for data interpretation and reporting.

% \input{contents/experiments/ablation_experiment}

% \section{Results}
% \input{contents/results/main_experiment}
% \input{contents/results/ablation_experiments}

In conclusion, this paper addresses the meaningful and challenging task of chart summarization by proposing ChartAdapter, a novel approach designed to bridge the cross-modal gap inherent in this task. ChartAdapter is designed for seamless integration with LLM-based methods, as demonstrated by its implementation within the state-of-the-art chart-to-text backbone, Qwen2VL-2B. Fine-tuning the entire framework using the newly developed large-scale ChartSumm dataset further enhances its capabilities. Comparative experiments against various existing methods, including state-of-the-art approaches, validate the effectiveness of the proposed ChartAdapter, highlighting its potential to advance the field of chart summarization significantly. 

Despite the progress we made, there are still further avenues for our research to explore, including 
\begin{enumerate*}[label=\textit{(\alph*)}]
	\item training a robust and SOTA model from scratch, 
	\item seeking for more efficient model structure, and 
	\item applying for other modalities other than linguistic and visual contents
\end{enumerate*}.

% \section*{Acknowledgment}
% The preferred spelling of the word ``acknowledgment'' in America is without 
% an ``e'' after the ``g''. Avoid the stilted expression ``one of us (R. B. 
% G.) thanks $\ldots$''. Instead, try ``R. B. G. thanks$\ldots$''. Put sponsor 
% acknowledgments in the unnumbered footnote on the first page.

% \section*{References}
% \vspace{5pt}
\bibliographystyle{ieeetr}
\bibliography{refe.bib}

% \AtNextBibliography{\normalsize}
% \printbibliography

\end{document}